# High-density single-molecule maps reveal transient membrane receptor interactions within a dynamically varying environment.

Nicolas Mateos[1,*], Parijat Sil[2,*], Sankarshan Talluri[2], Carlo Manzo[3,4,#], Satyajit Mayor[2,#], Maria Garcia-Parajo[1,5,#].

[1]ICFO – Institut de Ciencies Fotoniques, The Barcelona Institute of Science and Technology, 08860 Castelldefels (Barcelona), Spain.

[2]National Centre for Biological Sciences, Bangalore, India

[3]Facultat de Ciències i Tecnologia, Universitat de Vic - Universitat Central de Catalunya, 08500 Vic, Spain

[4]Institut de Recerca i Innovació en Ciències de la Vida i de la Salut a la Catalunya Central (IRIS-CC), 08500 Vic, Spain

[5]ICREA, Pg. Lluís Companys 23, 08010 Barcelona, Spain.

* Equally contributing authors

# Equally corresponding authors

Carlo Manzo: carlo.manzo@uvic.cat

Satyajit Mayor: mayor@ncbs.res.in

Maria F. Garcia-Parajo: maria.garcia-parajo@icfo.eu

**Author Contributions:** N.M., P.S., and S. T. prepared the biological samples and performed the single molecule imaging experiments. N.M. developed and performed simulations to validate the HiDenMap methodology. N.M. developed and validated the Rivers pipeline. N.M. and C.M. developed data analysis tools and analyzed the data. C.M., S.M., and M.F.G.-P. supervised the research. All authors conceived and discussed the experiments, interpreted the data, and wrote the paper.

**Competing Interest Statement:** The authors declare no competing interests.

**Content of this file:** combined main manuscript and 5 figures

**Abstract:**

Over recent years, super-resolution and single-molecule imaging methods have delivered unprecedented details on the nanoscale organization and dynamics of individual molecules in different contexts. Yet, visualizing single-molecule processes in living cells with the required spatial and temporal resolution remains highly challenging. Here, we report on an analytical approach that extracts such information from live-cell single-molecule imaging at high-labeling densities using standard fluorescence probes. Our high-density-mapping (HiDenMap) methodology provides single-molecule nanometric localization accuracy together with millisecond temporal resolution over extended observation times, delivering multi-scale spatiotemporal data that report on the interaction of individual molecules with their dynamic environment. We validated HiDenMaps by simulations of Brownian trajectories in the presence of patterns that restrict free diffusion with different probabilities. We further generated and analyzed HiDenMaps from single-molecule images of transmembrane proteins having different interaction strengths to cortical actin, including the transmembrane receptor CD44. HiDenMaps uncovered a highly heterogenous and multi-scale spatiotemporal organization for all the proteins that interact with the actin cytoskeleton. Notably, CD44 alternated between periods of random diffusion and transient trapping, likely resulting from actin-dependent CD44 nanoclustering. Whereas receptor trapping was dynamic and lasted for hundreds of milliseconds, actin remodeling occurred at the timescale of tens of seconds, coordinating the assembly and disassembly of CD44 nanoclusters rich regions. Together, our data demonstrate the power of HiDenMaps to explore how individual molecules interact with and are organized by their environment in a dynamic fashion, expanding the scope of current standard single-molecule and super-resolution imaging methods.

**Introduction**

Understanding the spatiotemporal organization and interactions of individual molecules in living cells is crucial to elucidate their functions. A wealth of research over the last 20 years has focused on the development and application of different forms of single-molecule methods to resolve their nanoscale organization and dynamic interactions between molecules and/or with their local environment. Yet, techniques that simultaneously combine nanometric scales and millisecond temporal resolutions are still challenging, because of which spatial and temporal studies are typically conducted separately. Super-resolution single-molecule localization microscopy (SR-SMLM) techniques such as PALM (1, 2), STORM (3), GSDIM (4, 5), and DNA-PAINT (6, 7) amongst others, provide detailed nanoscale information on the spatial organization of molecules at high labeling densities. This is achieved by effectively reducing the labeling density at each given time point such that the distance between emitting probes is larger than the Abbe's diffraction limit. At these sparse labeling conditions, the center of mass position of each individual fluorophore can be determined with nanometer precision. However, as SR-SMLM techniques still suffer from poor temporal resolution, imaging is commonly performed on fixed cells precluding dynamic studies at relevant temporal scales. Moreover, most SR-SMLM methods require specific photo-switchable or photo-convertible fluorescent probes together with the use of specific buffers that are not compatible with live cell imaging and also challenge the acquisition of simultaneous multi-color imaging.

To gain access to temporal information, several fluorescence-based optical techniques have been developed over the past years (8–12). A well-known and simple-to-implement approach is based on the combination of live-cell single-molecule imaging of sparsely-labeled molecules with single-particle tracking (SPT) algorithms, typically referred to as SPT (12). In SPT, each molecule is localized with nanometric precision and the localization positions tracked over multiple frames are used to reconstruct trajectories, from which dynamic information from each individual molecule can be extracted as a function of space and time (12–14). This technique provides similar localization precision as SR-SMLM at an increased millisecond temporal resolution (12, 15) and is compatible with the use of small labeling probes and different labeling strategies, including genetically-encoded fluorescent proteins (12). However, standard SPT also suffers from limitations: to localize and track individual molecules accurately, only a small population of molecules are labeled, resulting in poor statistics and sparse spatial sampling. Importantly, since each individual trajectory maps only a small region of space (limited by the photobleaching of the fluorophore), correlating molecular diffusion to spatial features of the surrounding environment becomes a daunting task. The short length of the trajectories also reduces the temporal scales at which molecular processes can be monitored. Thus, although fast, transient molecular interactions can be mapped (12), the slower dynamics of the environment and its impact on molecular organization remain inaccessible.

More recently, high-density SPT (HD-SPT) has emerged as a method to increase the collection of statistical information as compared to standard SPT (16). HD-SPT uses a higher labeling density that is temporally reduced via stochastic blinking (sptPALM (17)) or fluorophore binding (uPAINT (18)) to allow suitable tracking conditions. This approach significantly increases the amount of collected data, minimizing in turn the number of experiments needed. Yet, HD-SPT requires specific fluorescent probes (18) and the modulation of the labeling density slows down the spatial sampling (18), so that faster molecular dynamics and/or transient interactions with the environment are challenging to record.

Here, we propose an analytical approach that extracts spatiotemporal information from live-cell single-molecule imaging experiments performed at high labeling density (sub-saturating conditions) using conventional fluorescence probes. As such, we combine the advantages of SR-SMLM and standard SPT: single molecule nanometric localization accuracy together with millisecond temporal resolution in living cells. Our approach goes beyond other methods that map spatial diffusivity (19–22) since it quantifies relevant parameters of the spatiotemporal organization of proteins and their interactions with the surrounding environment. Importantly, by extending the observation times well beyond that of standard SPT, we unveil different temporal and spatial scales associated with transient interactions of molecules with a dynamically varying environment.

**Results and Discussion**

**Generation of high-density single-molecule maps (HiDenMaps).**

In our approach, the biomolecule of interest is labeled at high-density conditions, typically two orders of magnitude higher than for standard SPT, but at sub-saturating conditions so that individual molecules are resolved in each diffraction-limited image. At each video frame, we localize single molecules with nanometer precision. However, in contrast to standard SPT where the positions of individual molecules are reconnected in time to build up trajectories, here we simply collapse all the recorded localizations into a single image. The resulting high-density map (HiDenMap) thus contains all the localization positions of individual molecules as they dynamically explore the space (Fig. 1A). While the localization precision of a HiDenMap is comparable to SR-SMLM, the temporal resolution is given by the camera frame rate, typically tens of milliseconds. Under sufficiently high labeling conditions, a HiDenMap is visually similar to an SR-SMLM image (Fig. 1B, C), but acquired in living cells and in only a few seconds. Importantly, the HiDenMap encodes both spatial and temporal information that reports on the motion of molecules and their interactions with the environment.

Generating a HiDenMap is straightforward both experimentally and analytically. Nevertheless, for a faithful representation of the spatiotemporal maps, two main requirements should be fulfilled: imaging as many molecules as possible and for an extended observation time so that molecules explore the full space. These requirements depend in turn on different parameters such as the labeling density ($\rho$), the diffusion coefficient of the molecules ($D$), and the camera frame rate. To assess the influence of these parameters on the reconstruction of HiDenMaps, we performed simulations of Brownian particles with different diffusion coefficients, labeling densities, recording frame rates, and varying observation times (See Materials and Methods, Fig. S1 and *SI Text 1*). As expected, the higher the labeling density (with an upper value of ~1 molecule/$\mu m^2$, corresponding to the highest density at which single molecules can be readily localized within the diffraction limit) and the higher the frame rate, the shorter the observation time needed for molecules to explore the full space (Fig. S1A-C). In addition, the observation time reduces with larger diffusion coefficients (Fig. S1D). For a characteristic $D = 1$ $\mu m^2/s$, $\rho = 1$ molecule/$\mu m^2$, and camera frame rate of 100 Hz, the observation time required to map 90% of the space explored by individual molecules is less than 100 s (*SI Text 1*). These observation times are below the characteristic photobleaching time of most fluorophores used for single-molecule imaging (23, 24), indicating that faithful HiDenMaps of most proteins can be generated under realistic experimental conditions.

To prove the experimental feasibility of HiDenMaps, we first imaged supported lipid bilayers (SLB) containing Nickel-chelating lipids to anchor a decaHis-tagged SNAP-containing recombinant protein (His-SNAP-EzRA) labeled with JF549-SNAP. This system serves as a control for Brownian diffusing proteins with a diffusion coefficient $D \sim 1$ $\mu m^2/s$ (25). The HiDenMap rendered a homogeneous map (Fig 1D, 1$^{st}$ and 2$^{nd}$ columns) consistent with Brownian diffusion and simulations´ results. Indeed, at 75 s observation time with a camera frame rate of 100 Hz, we reconstructed the full space explored by the protein in the bilayer, indicating that our experimental settings are suitable for generating HiDenMaps that realistically capture the full spatial exploration of the diffusing molecules. We then performed single molecule imaging experiments of the transmembrane receptor CD44 expressed on live immature dendritic cells (imDCs) derived from primary monocytes. CD44 is a type I transmembrane adhesion receptor that can interact both with the intracellular cortical actin and with multiple molecules of the extracellular matrix (26). To establish our experimental settings, we first performed simulations of Brownian particles with a similar diffusion coefficient as CD44 ($D \sim 0.08$ $\mu m^2/s$) (27). The corresponding synthetic HiDenMap (Fig. 1D, 3$^{rd}$ column) showed a homogeneous distribution of localizations covering the full space at 100 s observation time with a camera frame rate of 33 Hz. Remarkably, the experimental HiDenMaps of CD44 significantly differed from the simulations of Brownian motion, revealing spatial features that are built over time, regardless of the frame rate and the observation times used (Fig. 1D, 4$^{th}$ and 5$^{th}$ columns). Moreover, whereas for SLBs, increasing the frame rate results in a homogenous enrichment of localizations in the HiDenMap, for CD44, increasing the frame rate essentially oversamples the previously built spatial features. These results clearly indicate that CD44 diffusion is nonhomogeneous and suggest that CD44 strongly interacts with its surrounding environment at the nano- (< 100 nm) and meso-scale (~ 1 μm), describing spatial patterns as the receptor dynamically explores the plasma membrane. As a whole, these experiments validate HiDenMaps as a methodology that provides a faithful representation of the space dynamically explored by individual molecules. Most importantly, they reveal marked differences between randomly diffusing molecules and those that interact with their environment.

**Pattern recognition algorithm to robustly quantify HiDenMaps.**

Although the quantification of HiDenMaps for molecules randomly exploring the environment is rather trivial since the maps are spatially homogeneous, mining information from heterogeneous HiDenMaps such as those observed for CD44 is highly involved. This heterogeneity is produced by non-Brownian dynamics and, likely, transient interactions of molecules with their environment, a feature that is common to most receptors in the cell membrane (12, 28). To extract spatiotemporal information from this type of HiDenMaps, we developed a pipeline for automatic pattern reconstruction. Our pipeline, named *Rivers*, combines the watershed (29, 30) and the skeletonization (31, 32) algorithms to extract the distinctive spatial features exhibited by the HiDenMaps. The workflow of the *Rivers* is depicted in Figure 2A. The first step of the pipeline takes all the raw localizations from the HiDenMap and removes those localizations that do not contribute to the spatial pattern, i.e., random localizations from Brownian motion. For this, we used a $1^{st}$-rank Voronoi density filter that classifies localizations into sparse (i.e., localizations from freely diffusing molecules) or dense (i.e., localizations composing the pattern) by considering the distance to the closest neighbors (see Materials and Methods for a detailed explanation). This step thus filters out the sparse localizations and retains the high-density localizations. Next, the pipeline generates a density map of the high-density localizations by defining a pixel of 50 nm (being half of the imaging pixel size) and by counting the number of localizations per pixel. Once the density image is generated, we apply two morphological operations in parallel and convolute their results. We first apply a watershed algorithm and connect the centers of high-density regions (29). However, as the watershed also generates a pattern on those regions of the image that are of lower density, we complement the watershed with a second morphological operation, namely skeletonization, to identify the skeleton of all high-density regions (32). Unfortunately, this algorithm also generates artifactual smaller ramifications within the high-density regions. By convoluting both the watershed and the skeletonization, we remove the artefactual links between lower-density regions generated from the watershed and the unwanted ramifications of the skeletonization. The final output is a binary image that contains the backbones of the high-density regions of the HiDenMap, i.e., the *in silico* reconstructed network pattern.

To validate the performance of the *Rivers* pipeline, we first turned to computer simulations. We generated an artificial network pattern (see Fig. 2B, 1st column) and performed simulations of molecules freely diffusing (Brownian motion with $D \sim 0.08 \; \mu m^2/s$, similar to that of CD44) without interaction with the network (Fig. 2B, top row), or undergoing interactions with the imposed network. Upon encountering the network, molecules transiently bind with probability $P_{inter}$. In the limit case of $P_{inter} = 1$, molecules remain permanently bound to the network (Fig. 2B, bottom row). The synthetic HiDenMaps (Fig. 2B, 2nd column) exhibit the expected randomly spread localizations for the case of $P_{inter} = 0$ (no binding), and a large number of localizations spatially coinciding with the imposed network for the case of $P_{inter} = 1$. We then used all the localizations from the synthetic HiDenMaps as input to the *Rivers* and extracted the final reconstructed network for both scenarios (Fig. 2B, 3rd column). For $P_{inter} = 0$, the *Rivers* pipeline reconstructs very small features (Fig. 2B, 3th column, top row) without meaningful overlap with the originally imposed network (Fig. 2B, 4th column, top row). Notably, for $P_{inter} = 1$, *Rivers* retrieves with extremely high fidelity the artificially imposed network (Fig. 2B, 4th column, bottom row).

We further performed simulations of diffusing particles for different values of $P_{inter}$ and used multiple indicators to quantitatively measure the similarities between the initially imposed and the *Rivers*-reconstructed networks (Manders overlap coefficient, Manders fractional coefficient, and Pearson coefficient) obtaining values above 80% for observations times ~ 100 s (see Fig. S2A-E and *SI text 2*). We also assessed the predictive capability of *Rivers* by computing the true positive and false positive rates as well as the receiver operating characteristic curve in a pixel-based manner (see Fig. S2F-H and *SI text 2*). Overall, our simulations indicate that *Rivers* performs remarkably well at retrieving and reconstructing heterogeneous patterns from HiDenMaps data.

We finally applied the *Rivers* pipeline to the experimental HiDenMaps of the labeled synthetic protein His-SNAP-EzRA in SLBs, as an example of Brownian motion, and of CD44 expressed on imDCs. Clearly, *Rivers* was unable to retrieve and reconstruct any discernable network on the lipid bilayer data (Fig. 2C), fully consistent with the Brownian diffusion of His-SNAP-EzRA, except for some isolated spots of accumulated localizations that most probably correspond to local defects on the bilayer that transiently trap the protein. In strong contrast, *Rivers* performed extremely well at reconstructing the spatial patterns observed on the experimental HiDenMaps of CD44 (Fig. 2C). These results thus validate the application of *Rivers* to reveal spatial features as molecules explore their surroundings.

**Extracting molecule dynamics at multiple temporal scales from HiDenMaps.**

HiDenMaps are built over time by the accumulation of the spatial localizations of single molecules as they diffuse in their environment. Therefore, segmenting the HiDenMaps into time windows should potentially reveal molecular dynamics at different timescales. To assess this capability in different scenarios, we first performed simulations on two extreme situations: the first one, when molecules diffuse in a purely Brownian fashion, and a second one, where the molecules remain permanently trapped upon encountering a static network. We then generated synthetic HiDenMaps by integrating the number of localizations over non-overlapping time windows of 5 seconds (Fig. 3A). As expected, in the case of Brownian motion, the HiDenMaps showed random exploration over time (Fig. 3A, first row), whereas in the case of permanent trapping, the temporal windows showed hotspots of accumulated localizations at the same spatial locations (Fig. 3A, second row). Notably, HiDenMaps retrieved for the experimental data on CD44 built over time windows of 5 seconds showed a mixture of both behaviors (Fig. 3A, third row). Indeed, we found scattered localizations arising from the random diffusion of the receptors as well as hotspots of accumulated localizations (pointed by arrows in Fig. 3A) consistent with trapping. Nevertheless, in contrast to the simulations of permanent trapping, the hotspots of localizations for CD44 were rather dynamic, having certain temporal persistence and eventually dissolving in time, or emerging at arbitrary time windows (see video S1). In addition, the hotspots displayed a directional movement as the localizations appeared to follow underlying paths. This is clearly visible when collapsing all the time windows in a final HiDenMap (last column in Fig. 3A, see also video S1): multiple hotspots of confinement built up in time into patterns in the HiDenMaps. These results thus suggest a complex dynamic behavior of CD44 on the plasma membrane with receptors alternating between Brownian diffusion and periods of transient trapping. The latter could originate from transient engagement of the receptors to a static underlying network, permanent receptor engagement to a network that is dynamically changing, or a combination of both.

To extract the temporal scales contained in the HiDenMaps, both inside and outside the underlying network, we developed an autocorrelation analysis ([Fig. 3B](#)). We first reconstruct the *Rivers* network from the experimental HiDenMap and use it as a mask to classify the HiDenMap localizations in two groups according to whether they reside inside or outside the network. After classification, both groups of localizations are separately used to build images with an overlapping sliding window of $\Delta t = 500$ ms (with steps of $\Delta t$ equal to the frame rate) and compute image autocorrelation curves as a function of time. In general, the curves were best fitted with a two-exponential function (see Materials and Methods) from which the characteristic decay times, $\tau_1$ and $\tau_2$, and their respective amplitudes $A_1$ and $A_2$ were extracted. Since the autocorrelation curves are generated for both groups of localizations, i.e., inside (in) and outside (out) the network, we obtain the following parameters: $\tau_{1,in}$, $\tau_{2,in}$, $A_{1,in}$, $A_{2,in}$ inside the network; and $\tau_{1,out}$, $\tau_{2,out}$, $A_{1,out}$, $A_{2,out}$ outside the network. This procedure thus allow us to fully distinguish different interaction dynamics of the molecules with their environment.

**Multiscale spatiotemporal interactions of actin-binding membrane receptors revealed by HiDenMaps.**

We next applied our set of quantitative tools to investigate the spatiotemporal organization of proteins present in the membrane of a cell. We examined CD44 endogenously expressed by imDCs and a set of synthetic proteins (expressed in CHO cells) that interact with the cortical actin with different strengths. In addition, and as an experimental control for Brownian motion and lack of any interaction, we also analyzed HiDenMaps of the His-SNAP-EzRA recombinant protein anchored to SLBs. For the synthetic proteins expressed on CHO cells, we particularly focused on a transmembrane actin-binding domain (TmEzABD) construct consisting of an extracellular folate receptor domain, a transmembrane I domain from the IgG receptor tagged to the actin-binding domain of Ezrin (Ez-ABD) as its cytosolic tail. Ezrin is a cytoplasmic protein belonging to the ERM family of proteins (Ezrin/radixin/moesin) (33, 34). ERM proteins function as linkers between transmembrane proteins such as CD44 and the actin cytoskeleton (35, 36). Thus, TmEzABD serves as a positive control for a receptor that constitutively binds to cortical actin. In addition, we also studied the TmEzABD* protein, which has the same domain structure as TmEzABD but with a point mutation on the actin-binding recognition domain, and thus only able to weakly interact with actin (37, 38). Finally, CD44 strongly binds to actin via association with Ezrin (26), but in addition, it also interacts with the extracellular matrix by binding to galectins and to other components of the extracellular milieu (26). Recent studies from our group already showed that CD44 mobility is highly influenced by cortical actin, exhibiting meshwork-like patterns on the cell membrane as the receptors dynamically explore and interact with their environment (39). Nevertheless, the temporal scales of these interactions have not been quantified.

We generated time-windows HiDenMaps for the three different receptors and observed a heterogenous exploration of the space by the TmEzABD and the TmEzABD* membrane proteins, visually similar to the patterns described by CD44. To extract their characteristic temporal scales we first performed the Voronoi filtering step to remove localizations that originate from random diffusion of the different proteins. To allow for a full comparison of the data, we also performed similar filtering on the SLB data as well as on simulations of Brownian diffusion to each set of experimental data (Fig. 4A). As expected, for the SLB data, the filter removes 99.5% of the localizations, comparable to the simulations of Brownian motion (Fig. 4A). When comparing the results of the three proteins (TmEzABD, TmEzABD* and CD44) to their corresponding Brownian motion simulations, the percentage of filtered localizations is much lower. This is because a significant number of localizations from the experimental data are identified by the Voronoi algorithm as belonging to high density regions, i.e., not random, and therefore, they are not removed by the filter (Fig. 4A). Although the differences between the three proteins are not statistically significant, the percentage of removed localizations is somewhat higher for TmEzABD* as compared to TmEzABD and CD44. These results are consistent with the fact that TmEzABD* interacts weakly with actin and therefore exhibits a higher degree of random motion. Interestingly, in the case of CD44, the spread on the percentage of removed random localizations is much larger than for the other two proteins most probably due to the complex interactions of CD44 with its environment.

We then reconstructed the corresponding *Rivers* networks from experimental HiDenMaps and used them as a mask to classify the HiDenMap localizations in two groups according to whether they reside inside or outside the network. We derived autocorrelation curves inside and outside the patterns for each of the proteins expressed in cells and extracted the characteristic temporal scales and their amplitudes, as described above (see Fig. S3). Regardless of the protein investigated, the autocorrelation curves for the localizations outside the patterns decayed fast consistent with a random exploration of the space (Fig. S3, blue curves). In contrast, much longer temporal decays were obtained for localizations inside the patterns (Fig. S3, orange curves), an indication of its temporal persistence. Note that for the SLB data, the Voronoi filter removed almost all the localizations consistent with the lack of an underlying pattern and therefore no *Rivers* network could be generated (see also Fig. 2C).

Figure 4B-D shows the fitted decay times and the amplitudes for the three receptors studied. Irrespective of whether the localizations were found inside or outside the patterns, we retrieved two characteristic times in the HiDenMaps, a fast decay ($\tau_1 \sim$ 2-4 s, Fig. 4B) and a rather slow decay ($\tau_2 \sim$ 10-40 s, Fig. 4C). In all cases, *outside* the patterns, $\tau_{1,out}$ was shorter ($\sim$ 2 s) than inside ($\tau_{1,in} \sim$ 4 s) and within the range of temporal scales expected for their Brownian diffusion (Fig. 4B). In addition, this fast component was dominant, since $A_{1,out} \sim$ 0.7-0.8 (Fig. 4D), implying that outside the patterns receptor diffusion is mostly Brownian. Interestingly, for the long decay component, $\tau_{2,out}$, the difference between CD44 and the two other proteins is remarkably large, i.e., $\sim$ 30 s for CD44, and $\sim$ 10 s for both TmEzABD and TmEzABD* (Fig. 4C). We interpret these results by considering that CD44 can interact with multiple components of the extracellular matrix including its ligand hyaluronan and galectins (26). These interactions could then be responsible for its longer $\tau_{2,out}$ value as compared to TmEzABD and TmEzABD*, that do not possess extracellular binding motifs, and therefore are less likely to interact with the extracellular matrix. Although treating the cells with lactose to reduce CD44 interaction with galectins did not change the overall $\tau_{2,out}$ values (Fig. 4C), it decreased the variability of the data, supporting our assignment that the long temporal component outside the patterns is related to the ability of CD44 to interact with the extracellular milieu.

When analyzing the dynamics *inside* the patterns, we found that the fast decay component $\tau_{1,in}$ for all receptors was longer than outside the network (Fig. 4B) and the $A_{1,in}$ values reduced with respect to $A_{1,out}$, i.e., $A_{1,in} \sim 0.4$-$0.5$ whereas $A_{1,out} \sim 0.7$-$0.8$ (Fig. 4D). The longer $\tau_{1,in}$ values together with their lower occurrence indicate that free diffusion of the receptors is hindered by the presence of an underlying network. Although the fast decay values were not statistically different amongst the proteins due to the large data variability, we found that TmEzABD exhibited longer $\tau_{1,in}$ as compared to the other proteins (Fig. 4B). Moreover, the median values of $\tau_{2,in}$ for TmEzABD and CD44 were longer ($\tau_{2,in} \sim 26$ s) than for the TmEzABD* construct ($\tau_{2,in} = 18$ s) (Fig. 4C). These results are fully consistent with the notion that the patterns described by the receptors as they diffuse on the cell membrane correspond to interactions with the cortical actin cytoskeleton, as we recently showed for CD44 (39). Since TmEzABD and CD44 strongly bind to actin, more persistent (i.e., longer) interaction times are thus expected as compared to the TmEzABD* construct whose interaction with actin is weak. Notably, the variability on the CD44 data is also large inside the network, an indication of the complex interactions of this receptor with its environment. Indeed, similar to the results obtained outside the network, treating the cells with lactose to reduce CD44 binding to galectins reduced the variability on $\tau_{2,in}$ and $A_{2,in}$, but without significantly affecting the median values.

It is challenging to associate these two characteristic times $\tau_1$ and $\tau_2$, to specific biological processes, since they could result from interactions of the receptors with their nano-environment, dynamic remodeling of the environment itself, or both. Nevertheless, we can envisage that the similar $\tau_{1,out}$ values obtained for all proteins outside the network likely correspond to the characteristic time for their Brownian diffusion. Indeed, considering the diffusion of CD44 to be 0.08 $\mu m^2/s$, the distance explored by the receptor during $\tau_{1,out} \sim 2$ s is $\sim 0.8$ $\mu m$, which is roughly the extent of the spatial dispersion of the single-molecule localizations observed on the HiDenMaps (see also Fig. 3A, bottom panels). Notably, $\tau_{1,in} > \tau_{1,out}$, particularly for TmEzABD, the construct that actively interacts with actin. Based on these results, it is highly plausible that the fast decay component inside the network corresponds to hindered diffusion of the receptors due to dynamic interactions with the actin cytoskeleton. Regarding the rather long decay component inside the network, the $\tau_{2,in}$ and $A_{2,in}$ values are quite similar for all the proteins, albeit slightly shorter for TmEzABD* and with larger $A_{2,in}$. Since TmEzABD* is less sensitive to changes in actin dynamics, the similarity of the $\tau_{2,in}$ and $A_{2,in}$ values for the other proteins suggests that the long $\tau_{2,in}$ component inside the network could correspond to the slow remodeling of the cortical actin cytoskeleton. Such long temporal remodeling is also in agreement with our recent results showing that the actin-dependent persistence of regions of CD44 confinement takes place at a similar temporal scale (39) (Fig. S4), and matches well the dynamics of actin reported by Lidke and co-workers (40). In summary, our analysis indicates that the fast decay component inside the network might result from dynamic interactions of receptors with the actin cytoskeleton while the long decay component arises from the slow remodeling of actin. Altogether, these results underscore the potential of HiDenMaps to reveal different spatiotemporal scales associated with the interactions of receptors in a dynamically changing environment.

**Linking nano- to meso-scale temporal scales using HiDenMaps combined with SPT.**

HiDenMaps are a versatile methodology that enables the correlation of the mesoscale spatiotemporal organization of molecules with single-molecule dynamics. This is because despite the high-labeling densities used to generate the HiDenMaps (typically 50 -100 nM) it is still possible to retrieve single-molecule trajectories from the same data by taking advantage of the progressive photobleaching occurring over time, decreasing the effective labeling density and enabling single molecule tracking.

We thus followed this approach to track individual CD44 molecules on CHO cells labeled with the SNAP-tag JF640 at a concentration of 100 nM. After progressive photobleaching, individual molecules could be tracked and superimposed to the HiDenMap built from localizations accumulated for 90 s at a camera frame rate of 60 Hz (Fig. 5A). We classified and/or segmented the trajectories according to whether the molecules had visited a region of high-density localizations (hotspot) or a low-density region of the HiDenMap (see

Materials and Methods and Fig. 5A) and analyzed different mobility parameters from multiple trajectories. Molecules diffused slower (Fig. 5B) and with a lower instantaneous velocity (Fig. S5A) in hotspot regions as compared to low-density localization regions. Moreover, whereas the turning angle between consecutive displacements was essentially isotropic in low-density localization regions, it clearly shifted towards 180º in hotspot regions, indicating receptor confinement (see Materials and Methods and Fig. S5B).

To better quantify these results, we identified transient confinement zones (TCZ) in single-molecule trajectories and classified the TCZ arrest times, $\tau_{TCZ}$ according to whether they occurred in hotspots ($\tau_{TCZ-HS}$) or low-density ($\tau_{TCZ-LD}$) localization regions of the HiDenMap (Fig. 5C). In addition, we performed simulations of Brownian motion and computed the $\tau_{TCZ}$ that could be detected with our TCZ algorithm, to compare with the experimental data. Notably, the distribution of $\tau_{TCZ-LD}$ was similar to that of the simulated Brownian motion trajectories likely resulting from a low occurrence of false-positive identifications. By contrast, the distribution of $\tau_{TCZ-HS}$ showed a tail towards longer confinement times as compared to that of $\tau_{TCZ-LD}$ and their occurrence was much higher (Fig. 5C). Thus, these results clearly show that individual molecules are preferentially arrested in the hotspot regions described by the HiDenMaps.

Finally, we computed the lifetime of the hotspots, i.e., temporal duration of the hotspots in the HiDenMaps before their disassembly (see Materials and Methods) and correlated them with the $\tau_{TCZ-HS}$ obtained from single trajectories (Fig. 5D). Notably, the lifetime of the hotspots is ~35 s, a value that is similar to the $\tau_{2,in}$ obtained from the autocorrelation analysis, confirming that the slow temporal component $\tau_{2,in}$ indeed corresponds to the dynamic remodeling of the network. Importantly, the confinement time of individual molecules in these hotspot regions is much shorter, i.e., $\tau_{TCZ-HS}$ ~ 250 ms, indicating that the arrest of molecules in these regions is much shorter and transient than the actual lifetime of the hotspots. Together, these results reveal that CD44 alternates between random diffusion and transient trapping by interacting with hotspot regions of the network and getting arrested for a few hundred milliseconds, while the network itself dynamically remodels in the tens of seconds time scale. As such, HiDenMaps combined with SPT reveals dynamic molecular interactions within a dynamically changing environment.

**Conclusions**

We have presented a new methodology based on the combination of a palette of analysis tools with live-cell single-molecule imaging at high labeling density (HiDenMaps). In stark contrast to standard single-molecule imaging and tracking methods that deliver information on the mobility of individual molecules on temporal scales at which the surrounding is essentially static, our approach provides access to different spatiotemporal scales involved in the interaction of individual molecules with their dynamic environment. Such a dynamic surrounding will not only impact on the mobility of molecules but also possibly on a large number of multi-molecular interactions. Moreover, by simply relying on the photobleaching properties of standard fluorescent probes, HiDenMaps can be readily combined with single-molecule tracking methods extending the range of temporal scales that can be accessed using the same set of experimental data.

The data shown here has been retrieved using a single-color scheme, but HiDenMaps can be extended to multi-color applications to study the correlation between multiple proteins simultaneously. Furthermore, by incorporating a second detection channel, HiDenMaps can be combined with other fluorescence-based approaches, including fluorescence resonance energy transfer (FRET) enabling the mapping of different properties of the plasma membrane. For example, in previous work from our groups using a less developed version of the analysis pipeline presented here, we mapped the interactions of CD44 with the cytoskeletal meshwork as well as simultaneously observing its nanoclustering behavior by means of homoFRET-anisotropy (39). Together with this level of organization, our new data suggest that the TCZs correspond to the regions rich in nanoclusters, and are likely to be a consequence of actin-induced nanoclustering. This approach has revealed the hierarchical landscape of CD44 at the cell surface at multiple scales and its interaction with the membrane environment.

We have exploited HiDenMaps to investigate the spatiotemporal organization of different proteins either anchored to SLBs or expressed on living cell membranes. Overall, our results showed that whereas on SLBs diffusion is essentially Brownian, membrane proteins in living cells exhibit a highly heterogenous diffusion at multiple spatiotemporal scales. Indeed, their mobility was highly affected by short transient interactions with cortical actin (~ 250 ms), but also importantly, by the dynamics of actin itself, on a much longer temporal scale (tens of seconds). The hotspots in turn are likely to represent the regions of the membrane where nanoclusters are enriched, outlining a membrane region of distinct composition and physical characteristics. Although we have focused on membrane receptors interacting with cortical actin, HiDenMaps can be equally applied to a multitude of other molecular interactions occurring on the cell membrane, or intracellularly (41).

In summary, the amount of information that can be obtained in a simple HiDenMap experiment is extensive, providing exquisite detail on how individual molecules sense and interact with their environment in a dynamic fashion. We envisage that this methodology will open new frontiers in image acquisition and analysis filling the gap in the biophysics field by overcoming the limitations of standard single-molecule tracking and super-resolution microscopy approaches.

**Materials and Methods**

**Cell culture and labeling conditions for high-density single molecule imaging of TmEzABD and TmEzABD*:** TrVb1 cells (Chinese hamster ovary, CHO cells expressing human transferrin receptor) stably expressing the Folat binding domain of the folate receptor (FR), FR-TmEzABD or FR-TmEzABD* were cultured in Ham's-F12 media (HF-12) (HiMedia, Mumbai, India), supplemented with 10% fetal bovine serum (FBS) (GibcoTM, 16000044), cocktail of Penicillin, Streptomycin, L-Glutamine (PSG) (Sigma, G1146-100ml), Geneticin (200ug/ml) and Hygromycin (100ug/ml) at 37 °C and 5% $CO_2$. For imaging, cells were seeded on glass coverslip fitted 35mm cell culture dishes and grown in Ham's-F12 media (with PSG but without selection antibiotics) for 2 days. Cells were labeled using MOV F(ab) fragment (a generous gift from Dr. Mariangela Figini, Dipartimento di Ricerca Applicata e Sviluppo Tecnologico (DRAST), Fondazione IRCCS Istituto Nazionale dei Tumori, Milan) was conjugated with STAR-635 (Abberior ST635). MOV F(ab) specifically recognizes and binds to the FR domain of the receptor. The labeling concentration was ~50-100nM. MOV-635 was diluted in 10% FBS containing phenol red free HF-12 and incubated at 37 °C, 5% CO$_2$ for 5 minutes. Cells were washed and imaged in HEPES buffer containing 2mg/ml glucose, in a stage-top incubator, set to 37 °C.

**Cell cultures and labeling conditions for high-density single molecule imaging of CD44.** Single-molecule imaging of CD44 was performed on different cell types. For the data shown in Figures 1-4, we used immature dendritic cells (imDCs) derived from peripheral blood mononuclear cells (PBMC). To generate imDCs, we first obtained PBMCs using with a Ficoll-Hypaque gradient (Alere Technologies AS) from HIV-1-seronegative donors, and monocyte populations (>97% CD14+) were isolated with CD14-positive selection magnetic beads (Miltenyi Biotec). imDCs were obtained after culturing the monocytes in complete RPMI supplemented with 1,000 IU/ml granulocyte-macrophage colony-stimulating factor (GM-CSF) and interleukin-4 (IL-4) (both from R&D) for seven days and replacing media and cytokines every two days. In experiments with lactose, cells were treated with 100 nM lactose from day 4. Experiments were conducted at day 6. For single molecule experiments in living cells, CD44 labeling on imDCs was performed using biotinylated single chains generated in-house from monoclonal mouse anti-human CD44 (clone G44-26, BD-Biosciences) conjugated with streptavidin QD-605 (Thermofisher), at a labeling concentration of 30-50 nM. For the data shown in Figure 5, we used CHO cells cultured as described above. Labeling of CD44 on CHO cells was performed using a SNAP-tag JF640 at a concentration of 100 nM.

**Supported Lipid Bilayers (SLB) composition.** No.1 Coverslips (CS 25R/ 64-0705, Warner Instruments), were sonicated with Hellmanex III (0.5%) for 40 min followed by 5M KOH for 10 min. The glass was washed extensively in MilliQ after each treatment. The glass was then dried under N$_2$ and stored in a desiccator until further use. Just before making the bilayer, a coverslip was taken out of the desiccator, and a cylindrical chamber (made from a cut PCR tube) was stuck onto the coverslip with transparent UV glue. The coverslip was then placed in an Ozone/UV cleaner for 15 min and washed afterwards with PBS. Small single unilamellar vesicles (SUV) with DOPC (98%) and NiNTA-DGS (2%) were prepared in advance according to the protocols mentioned in (25). 2uL of 4mM SUVs were added to the chambers and were incubated for 15 min. The unbound vesicles were washed off with buffer. The coverslip was incubated with 0.1mg/ml of beta-Casein for 10 min to block the exposed surfaces where vesicles did not form a bilayer which was subsequently washed off. His-SNAP-EzRA (a SNAP-tagged version of the construct HYE-R579A used in (25)) was used as a bilayer marker. His-SNAP-EzRA has a Deca-His tag which allows it to bind to the Nickel-containing lipids on the bilayer. It has a SNAP-tag for visualization, which was labeled with JF-549. His-SNAP-EzRA was added to the bilayer at a final concentration of 100 picomolar and allowed to bind for 40 min before washing it off. PCA/PCD/Trolox described in (42) were added at 4mM, 125nM, and 2mM respectively to reduce photobleaching and triplet state transitions.

**Sample preparation and super-resolution imaging of CD44 in fixed imDCs.** Day 6 imDCs were plated on an eight-well plate Lab-Teck #1 at a density of 50.000 cells/well. Cells were incubated at 37°C for 1hour before fixation with 4% PFA in PBS for 15 min at room temperature. After fixation, samples were blocked using 3% wt/vol bovine serum albumin (BSA) in PBS for 30 min at room temperature. Cells were labeled with primary antibody rabbit-antihuman CD44 (HPA005785, Sigma-Aldrich) at a concentration of 5 µg/ml for 1 hour at room temperature. The corresponding secondary antibody, an anti-rabbit antibody, was tagged in-house with Alexa Fluor 647 (Invitrogen) as a reporter and Alexa Fluor 405 as an activator. We incubated the secondary antibody for 1 hour at room temperature.

Super-resolution imaging was performed using a commercial Nikon Eclipse Ti system (N-STORM) with a 100x oil objective with NA 1.49 using TIRF illumination. The detector was an ANDOR technology EMCCD iXon 897 camera, with a 256x256 pixel ROI and a pixel size of 160 nm. Excitation of Alexa Fluor 647 was achieved using the 647 nm laser line from an Agilent Technologies laser box. The Nikon-developed NIS software was used for acquiring the data. In a pre-acquisition step, the sample was irradiated with the 647 nm laser at 70% (~107 mW) to force the majority of molecules into a long-lived dark state. Afterward, we acquired STORM data while maintaining the 647 nm laser at constant power and using low-power activator (405 nm) excitation, which we increased incrementally, from 0.5% to 100%, as the acquisition progressed. The STORM buffer used was Glox solution (40 mg/ml Catalase (Sigma Aldrich), 0.5 mg/ml glucose oxidase, 10% glucose in PBS) and MEA 10 mM (Cysteamine MEA (Sigma-Aldrich; #30070-50G) in 360 mM Tris-HCl). STORM images were analyzed using a DBSCAN algorithm (43) to determine the clusters of localizations, with an epsilon of 50 nm and 50 as the minimum number of points. Once the clusters were identified we computed their radius.

**Single-molecule imaging on supported lipid bilayers.** The SLBs were imaged on a Nikon Ti-E microscope body equipped with a motorized total internal reflection (TIRF) arm (from Nikon) with a 100x, 1.45NA objective. Agilent (MLC-300) laser combiner was used with a 561 nm laser line, giving 9 mW at the back focal plane. Images were acquired on Prime95B sCMOS camera at a frame rate of 100 Hz for 2 min.

**High-density single-molecule imaging of different membrane receptors.** The experiments were performed on two different single-molecule setups. Imaging of CD44 on CHO cells was performed on a home-built TIRF microscope consisting of a Nikon Eclipse Ti body and a Nikon 100X Apochromat 1.49 NA objective. The set-up is equipped with a high-speed CMOS camera (FASTCAM-SA1, Photron, Tokyo – Japan – (44–46)) coupled to a micro-channel plate intensifier (C8600-03, Hamamatsu Photonics, Hamamatsu, Japan) by way of an optical-fiber bundle. Imaging was performed at a frame rate of 60 Hz and an acquisition time of 90 seconds. The excitation power ($\lambda = 640$ nm) at the back focal plane was of 18 mW. Imaging of CD44 on imDCs was performed on a home-built TIRF microscope equipped with a Nikon CFI APO TIRF 60x, 1.45 NA objective. Excitation was provided by a solid-state 488 nm laser line and the fluorescence emission was collected using an EM-CCD camera from Hamamatsu. Imaging was performed at a frame rate of 30 Hz and an acquisition time of 90 seconds.

**Simulations of Brownian molecules.** We performed stochastic simulations of Brownian molecules with diffusion coefficient *D* in 2D. We defined our imaging region to be a square of side L=256 pixels (100 nm per pixel) and to avoid edge effects, we randomly placed 1500 molecules in a 256-by-256 pixels grid being the center of our imaging region. The total observation time was set to 1000 s and the frame rate to 100 Hz. For the diffusion, at each temporal step molecules were displaced according to a normal distribution $\delta x, y \sim N(0, \sqrt{2 \cdot dt \cdot D}\ )$. While diffusing, molecules could not interact with each other. In the case of results showing slower frame rates than 100 Hz, we down-sampled the data from the 100 Hz. With this, we were able of generating simulations at multiple frame rates from a fast simulation.

**Technical requirements for the generation of HiDenMaps that cover the full space explored by molecules.** We investigated how labeling density, frame rate, observation time, and diffusion coefficient affect the HiDenMaps. To minimize the computational cost of the simulations, we performed five simulations per each diffusion coefficient ($D$ = [0.01, 0.1, 1, 10] μm$^2$/s) at high density ($\rho$ = 1.22 molecules/μm$^2$) and fast frame rate (100 Hz). For the simulations with lower labeling density, we down-sampled the trajectories from the high-density simulations. For slower frame rates we followed a similar approach by removing frames. Figure S1 and *SI Text 1* summarize the results for the requirements of HiDenMaps.

**Simulations of diffusing molecules interacting with an imposed network.** These simulations add a layer of complexity to the simulations of Brownian diffusion. We first generated an imposed network consisting of a Voronoi network with 0.5 μm$^2$ mean patch size and a filament thickness of 10 nm (similar to actin). We defined space, number of diffusing molecules, observation time, and frame rate in the same way as for the simulations of Brownian motion. However, in the current simulations, molecules diffuse randomly (Brownian motion with $\delta x, y \sim N(0, \sqrt{2 \cdot dt \cdot D})$) within the free spaces of the imposed network. When a step $\delta x, y$ crosses a filament or ends on the filament, then the molecule has a certain probability, $0 < P_{inter} \leq 1$, to interact with the network. The interaction is determined by randomly generating a number from a uniform distribution between 0 and 1. If the number is smaller than $P_{inter}$, then the interaction takes place, otherwise the molecule proceeds with its Brownian motion. The interaction with the network consists of staying on the network for a time that is given by an exponential probability distribution with $\lambda = \tau_{inter}$. After this time, the molecule can rebind (interact again) to the network with a probability, $0 < P_{rebind} \leq 1$. If the molecule rebinds to the network, then it will spend some time in the network, as before. For our simulations, we studied two scenarios: $P_{rebind} = 0$, or rebinding with $P_{rebind} = P_{inter}$. We iterate this process for all the simulated molecules for a total observation time of 1000 s, and frame rate of 100 Hz. Finally, once all the trajectories have been generated, we add white Gaussian noise to all localizations with a standard deviation of 20 nm to account for a finite localization accuracy.

**Voronoi tessellation-based filtering.** We adapted a 1st rank Voronoi tessellation algorithm based on previously published work by Levet et al (47, 48) in order to filter out localizations from the HiDenMap that arise from Brownian motion. The algorithm consists of two parts: first, we compute the normalized rank 1 Voronoi density for each localization and second, we apply a filter based on these densities. Given a set of N seeds, $s_{k,k\in[1,N]}$, we perform a Voronoi tessellation of the space such that each seed, $s_i$, belongs within a polygon, $P_i$, of area $A_i$. The 1st rank neighbors of seed $s_i$ are those seeds whose polygons share an edge with seed $s_i$.

The rank 1 Voronoi density is computed as follows:

$$\delta_i^{1^{st}} = \frac{1+n_i^1}{A_i+\sum A_{i,j}^1} \quad (1)$$

Where $n_i^1$ corresponds to the number of 1st rank neighbors, $A_i$ the area of the Voronoi cell for seed $s_i$ and $A_{i,j}^1$ are the Voronoi areas of the 1st rank neighbors. We then normalized the density by dividing $\delta_i^{1^{st}}$ by the mean density of uniformly distributed localizations (same number as the experimental data):

$$\hat{\delta}_i^{1^{st}} = \frac{\delta_i^{1^{st}}}{\delta_{rand}} \quad (2)$$

The normalization is performed in order to center the distribution around 0 in a logarithmic scale. Figure S6A shows an example on how the normalized 1st rank Voronoi density is computed for the seed in the purple Voronoi cell.

In order to set a threshold to determine the cut-off between high-density and low-density localizations, we generated a set of simulations of Brownian diffusion with different diffusion coefficients ($D = 10^{-3}, 10^{-2}, 10^{-1}, 10^0$ and $4.8 \cdot 10^{-2}$ $\mu m^2/s$). Figure S6B shows the distributions of the normalized 1st rank Voronoi densities in a logarithmic scale. For slow diffusions, $D < 10^{-2}$ $\mu m^2/s$, the distributions are shifted towards the right (high densities) since the molecules do not explore homogenously the space and the distance between consecutive steps is small. Nevertheless, for $D > 10^{-2}$ $\mu m^2/s$, the distributions are centered around zero, similar to randomly distributed localizations. To set the threshold, we computed the cumulative density function (CDF) for the normalized 1st rank Voronoi density (Fig. S6C) and screened the filtering efficiency by testing thresholds on the uniformly distributed localizations from 90% up to 99.99% (Fig. S6D). From this plot, it is clear that we need to define the threshold as the 99.99% of the CDF for the normalized 1st-rank Voronoi density.

***In silico* validation of the *Rivers* pipeline**. To validate the performance of the *Rivers* pipeline, we performed simulations of molecules interacting with an artificially imposed network with different interacting probabilities ($P_{inter} = [0, 0.25, 0.5, 0.75, 1]$) without rebinding ($P_{rebind} = 0$) and with an interaction time, $\tau_{inter} = 0.5\ s$. Note that the simulation of $P_{inter} = 0$ corresponds to purely Brownian motion. The simulations were performed for $D = 0.05\ \mu m^2/s$, comparable to that of CD44.

We reconstructed the *Rivers* network at different observation times for the HiDenMap ($T_{obs}$ = 20, 40, 60, … 300 s) and computed the Pearson coefficient, the Mander's overlap coefficient (MOC), and the Mander's fractional coefficient (MFC) between the reconstructed *Rivers* network (R) and the imposed network (N) using the following equations:

$$MOC = \frac{\sum_{i,j}(R_{i,j} * N_{i,j})}{\sqrt{\left(\sum_{i,j} R_{i,j}^2\right) \cdot \left(\sum_{i,j} N_{i,j}^2\right)}} \quad (3)$$

$$MFC = \frac{\sum R_N}{\sum R} \quad (4)$$

where $R_N$ refers to the pixels of R that coincide with full pixels in N. In addition, we also quantified the false positive rate (FPR), the true positive rate (TPR) and the Receiver Operating Characteristic (RoC) curve in a pixel-based manner.. The results are included and discussed in detail in Figure S2 and *SI Text 2*.

**Spatiotemporal autocorrelation analysis of HiDenMaps**. We first reconstructed the *Rivers* network from the HiDenMap of each receptor and used it as a mask to classify the localizations of the HiDenMap in two populations: localizations falling inside the network and outside the network. We then accumulated localizations in time windows of Δt = 500 ms for both populations. The length of the window results from a trade-off between accumulating a sufficient number of localizations to reveal spatial patterns and retaining a high temporal resolution. We further computed the autocorrelation curve as follows:

$$G_0(t_{lag} = m\Delta t) = \frac{1}{N-m} \sum_{i=1}^{N-m} \frac{<I(\Delta t_i)*I(\Delta t_i+m\Delta t)>}{<I(\Delta t_i)>\cdot<I(\Delta t_i+m\Delta t)>} \quad (5)$$

Here, $I(\Delta t_i)$ refers to the image of the i-th temporal window. We normalized the curves to the first point ($t_{lag}$ = 0 s) and finally fitted the decay curves from the second point onwards until $t_{lag}$ = 50 s. For the fitting we tried several components, finding that the best fitting to the experimental data was obtained using a two-exponential decay with a constant term:

$$F(t) = \tilde{A}_1 e^{-t/\tau_1} + \tilde{A}_2 e^{-t/\tau_2} + B \quad (6)$$

We performed the fitting using the MATLAB's *Curve Fitting Tool,* setting the bounds of $\tilde{A}_1, \tilde{A}_2,$ and $B \in [0,1]$ and $\tau_1$ and $\tau_2 \in [0, \infty)$. We chose '0.5' as the starting point for all the variables. The rest of the parameters used for the fittings are found in Table S1. Finally, once the fitting was performed, we rescaled the amplitudes of the exponential decays as follows:

$$A_1 = \frac{\tilde{A}_1}{\tilde{A}_1+\tilde{A}_2} \quad \& \quad A_2 = \frac{\tilde{A}_2}{\tilde{A}_1+\tilde{A}_2} \quad (7)$$

**Generation of single-molecule tracking trajectories.** We used the ImageJ plugin Trackmate for the detection of the single molecules (Diameter blob = 8 pixels and Threshold = 1.5). For the tracking, we used the simple LAP tracker with a maximum distance between frames of 5 pixels and we did not allow for missing frames from blinking Overall, we obtained >1000 trajectories for N= 9 cells.

**Trajectory segmentation using simultaneously obtained HiDenMaps.** We segmented the trajectories according to whether the molecule had visited a hotspot region of high-density of accumulated localizations or a region with low-density of localizations of the HiDenMaps. For this, we first defined a 1-by-1 $\mu m^2$ region of interest (ROI) around the trajectory. To establish if a region of the ROI has a high- or a low density of localizations, we performed a 1[st]-rank Voronoi density segmentation (taking the 90% of the CDF for uniformly distributed localizations as thresholds). To then segment the trajectory, we proceed as follows: for each localization of the trajectory, we define a circle of 50 nm in diameter and count how many high- and low-density localizations of the HiDenMap fall within the circle. We then compute the ratio of high-density localizations vs the total number of localizations within the circle. If the ratio is larger than 50%, then we annotate the localization in the trajectory to be on a hotspot region, otherwise it is annotated as falling on a low-density region. Finally, to split the trajectory according to high-density and low-density regions, we consider that at least three consecutive frames must be in the same type of region to classify that given segment. For each trajectory, we stitch together all the partial segments to obtain, in the best case, two segments per trajectory: one coinciding with the hotspot regions and the second one belonging to the low-density localizations region.

**Quantification of segmented trajectories.** Once the trajectories have been classified according to the regions from the HiDenMap they visit (i.e., hotspots or low-density localization regions) we quantified the segments by means of the apparent diffusion coefficient, $D_{1-4}$, (for segments longer than 20 frames), the instantaneous velocity, and the turning angle between consecutive segments as described in Refs. (12, 41).

**Transient confinement zones (TCZ) and identification of hotspot regions.** To identify TCZ of individual trajectories we used an algorithm adapted from Simson et al (49) and measured the lifetime of the TCZs. To correlate the TCZs with confinement regions of the HiDenMap close to the TCZ, we proceeded as follows: for each trajectory with a TCZ, we determined an ROI around it (± 200 nm) on the HiDenMap and searched for high-density localization regions (hotspots) using the 1$^{st}$ rank Voronoi-based tessellation. Then, if a hotspot exists, we calculated its lifetime before disassembly by measuring the time it takes for the cumulative distribution function of the localization appearances in the hotspot to increase from 5% to 80%.

**Acknowledgments:**

The research leading to these results has received funding from the European Commission H2020 Program under grant agreement ERC Adv788546 (NANO-MEMEC) (to M.F.G.-P.), Government of Spain (Severo Ochoa CEX2019-000910-S, State Research Agency (AEI) (PID2020-113068RB-I00 / 10.13039/501100011033 (to M.F.G.-P.), Fundació CELLEX (Barcelona), Fundació Mir-Puig and the Generalitat de Catalunya through the CERCA program and AGAUR (Grant No. 2021 SGR01450 M.F.G.-P.). N.M. acknowledges funding from the European Union H2020 under the Marie Sklodowska-Curie grant 754558-PREBIST. C.M. acknowledges support through grant RYC-2015-17896 funded by MCIN/AEI/10.13039/501100011033 and "ESF Investing in your future", grants BFU2017-85693-R and PID2021-125386NB-I00 funded by MCIN/AEI/10.13039/501100011033/ and FEDER "ERDF A way of making Europe", and grant AGAUR 2017SGR940 funded by the Generalitat de Catalunya. S.M. acknowledges a JC Bose National Fellowship from the Department of Science and Technology (Government of India), support from India Alliance Margadarshi Fellowship (IA/M/15/1/502018), and support from the Department of Atomic Energy (Government of India) under Project RTI 4006 to NCBS.

**Figure legends**

**Figure 1: Methodology for the generation of HiDenMaps and requirements.** (A) Schematic on how HiDenMaps are generated. Individual fluorescent molecules are localized at each frame of the video and collapsed into a single HiDenMap image. Scale bar is 10 μm. (B) Comparative SR-SMLM image obtained with STORM (top panel) and HiDenMap (bottom panel) of the transmembrane receptor CD44 expressed on immature dendritic cells (imDCs). The STORM image has been obtained on fixed imDCs while the HiDenMap image has been generated by integrating all the single-molecule localizations during 1 second video recorded at a frame rate of 30 Hz on living imDCs. Scale bar is 2 μm. (C) Radius of clustered localizations determined by DBSCAN for STORM (dark grey) and HiDenMap (light grey) images (see Materials & Methods). Median cluster radius for STORM and HiDenMap images are 73 nm and 71 nm, respectively. STORM images from 8 different cells and HiDenMap images from 11 different cells have been analyzed. (D) HiDenMaps at multiple observation times (vertical) and different frame rates (horizontal) for experimental data on supported lipid bilayers ($1^{st}$ and $2^{nd}$ columns), Brownian motion simulations of CD44 ($3^{rd}$ column), and experimental data of CD44 on imDCs ($4^{th}$ and $5^{th}$ columns). Scale bar is 1 μm.

**Figure 2. Pattern recognition algorithm to robustly quantify HiDenMaps.** (A) Schematic illustration of the *Rivers* pipeline to generate *in silico* reconstructed patterns from experimental HiDenMaps (see main text for details). (B) Simulated examples for a molecule exhibiting two types of diffusing behavior (Brownian motion without interaction with an underlying network, $P_{inter} = 0$, first row; and interacting with a network with $P_{inter} = 1$, second row). The $1^{st}$ column shows the imposed simulated network. The $2^{nd}$ column corresponds to the resulting HiDenMap image and the $3^{rd}$ column corresponds to the *in silico* reconstructed pattern obtained from *Rivers*. The overlay of all the columns is shown in the $4^{th}$ column, with white color depicting the overlap between the *Rivers*-reconstructed network and the originally imposed one. Scale bar is 2 μm. (C) HiDenMaps ($1^{st}$ and $3^{rd}$ columns) and the *Rivers*-reconstructed pattern ($2^{nd}$ and $4^{th}$ columns) for supported lipid

bilayers (1st and 2nd columns) and CD44 (3rd and 4th columns). The color-coding in the HiDenMaps corresponds to the number of localizations per pixel. Scale bar is 2 μm.

**Figure 3. Extracting dynamic information from HiDenMaps using autocorrelation analysis.** (A) HiDenMaps generated by accumulating single molecule localizations in time windows of 5 seconds. Three arbitrary time windows are shown: $T\in[20,25]$ $s$ (1st column), $T\in[55,60]s$ (2nd column) and $T\in[90,95]s$ (3rd column). The last column shows the overlay of the three time-windows. The different rows correspond to simulations of Brownian motion (1st row), simulations of permanent trapping on the network (2nd row), and experimental data obtained for CD44 (3rd row). The white arrows on the CD44 data highlight hotspots of accumulated localizations. The number of localizations at each time window is the same for the three scenarios. Scale bar is 2 μm. (B) Schematic illustration showing the analysis pipeline for the autocorrelation analysis. We used *Rivers* to generate a mask of the underlying network and used it to classify HiDenMap localizations falling inside or outside the network. For each set of localizations, we performed image correlation analysis and extracted the characteristic temporal scales (see main text for details).

**Figure 4. Multiscale spatiotemporal interactions of actin-binding membrane receptors revealed by HiDenMaps.** (A) Percentage of random localizations filtered out after the Voronoi filtering step, for the experimental data (light grey) and the corresponding simulations of Brownian motion (dark grey). (B-D) Boxplots for the parameters obtained from fitting the autocorrelation decay outside (light blue) and inside (light orange) the patterns using a two-exponential function, for the different membrane receptors. (B) The fast temporal decay, $\tau_1$. (C) The slow temporal decay, $\tau_2$. (D) The normalized amplitudes, $A_1$ (left axis) and $A_2$ (right axis). Number of cells (and ROIs analyzed per receptor): TmEzABD*: 12 (37); TmEzABD: 6 (16); CD44: 14 (44); and CD44 after lactose treatment: 15 (72). The statistical test performed is the non-parametric Kruskal Wallis.

**Figure 5. Linking nano- to meso-scale temporal scales by correlative HiDenMap and SPT.** (A) Single-molecule trajectories of CD44 (in magenta and pointed with white arrows) overlaid on a simultaneously obtained HiDenMap (green-to-yellow, i.e., low to high-density of localizations, respectively). The panel on the right corresponds to an enlarged region of the HiDenMap with an imposed single-molecule trajectory, segmented in two

parts: white line for the segment being confined in a HiDenMap hotspot, and black line for the segment diffusing freely in the low-density localization regions of the HiDenMap. Scale bar in the main image is 2 µm. Scale bar in the enlarged image is 200 nm. (B) Distribution of the instantaneous diffusion coefficient ($D_{1-4}$) for the segmented trajectories visiting hotspots (light grey) or low-density localization regions (dark grey). (C) TCZ times for CD44 trajectories in low-density localization regions, $\tau_{TCZ\text{-}LD}$ (dark grey), and in hotspots regions, $\tau_{TCZ\text{-}HS}$ (light grey). TCZ times detected on simulated trajectories of Brownian motion are shown in white. Number of TCZs detected: 840 and 2904 for low-density localization and hotspot regions, respectively. (D) 2D histogram correlating the TCZ times in the hotpots regions, $\tau_{TCZ\text{-}HS}$ (x-axis) with overlapping hotspot lifetimes (y-axis). The distributions of $\tau_{TCZ\text{-}HS}$ and hotspot lifetimes are shown on the top and right sides of the 2D histogram, respectively.

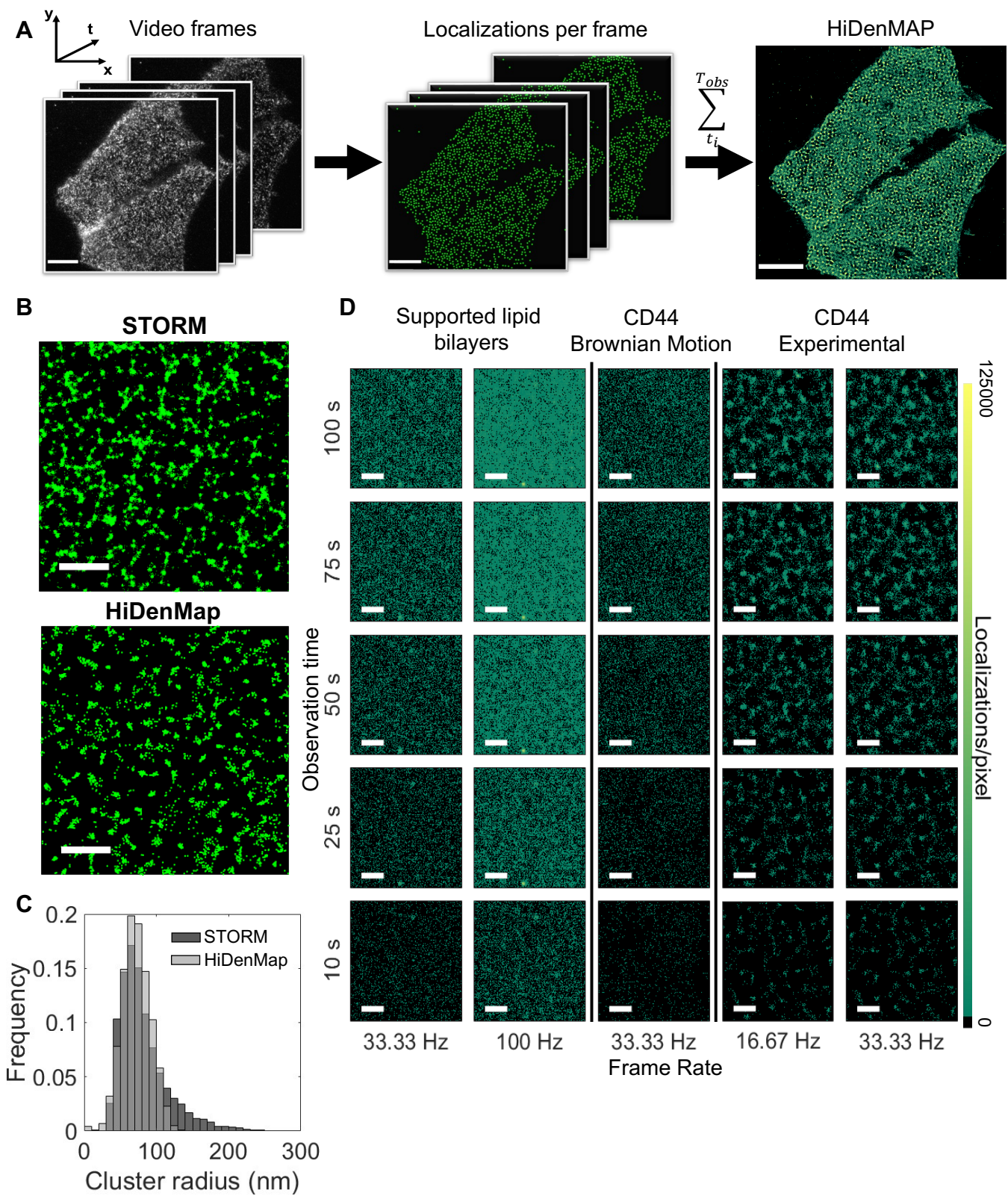

Figure 1

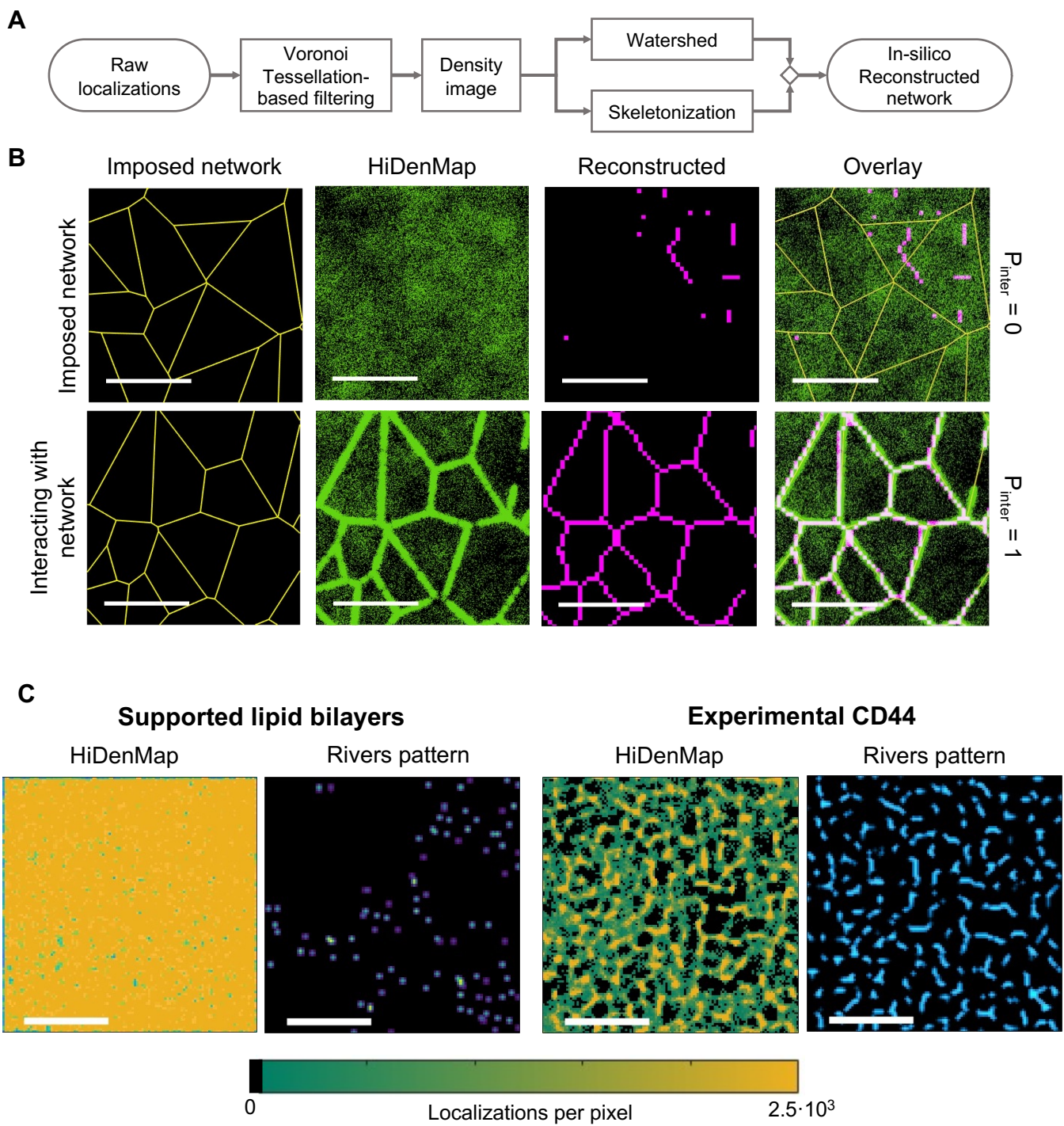

Figure 2

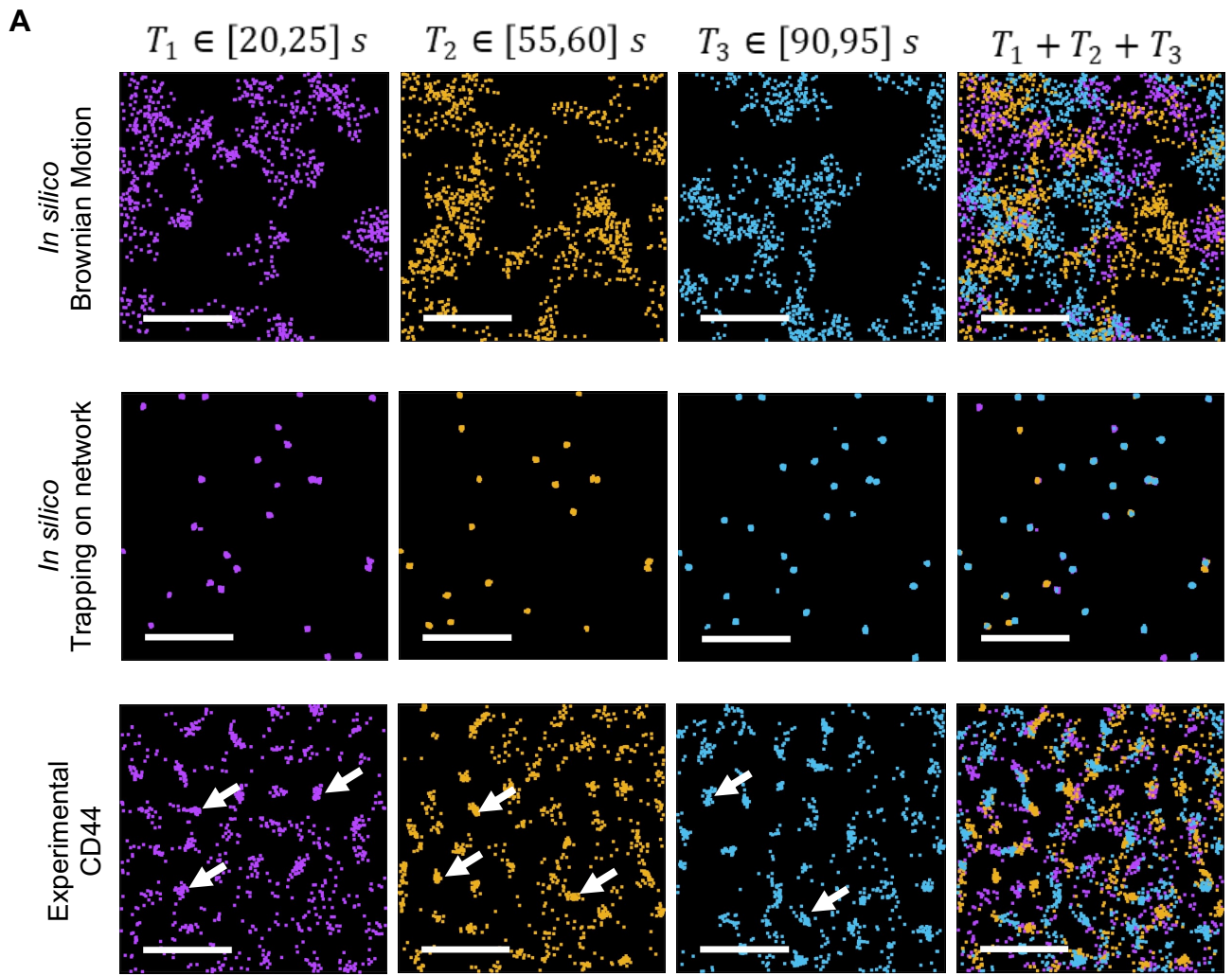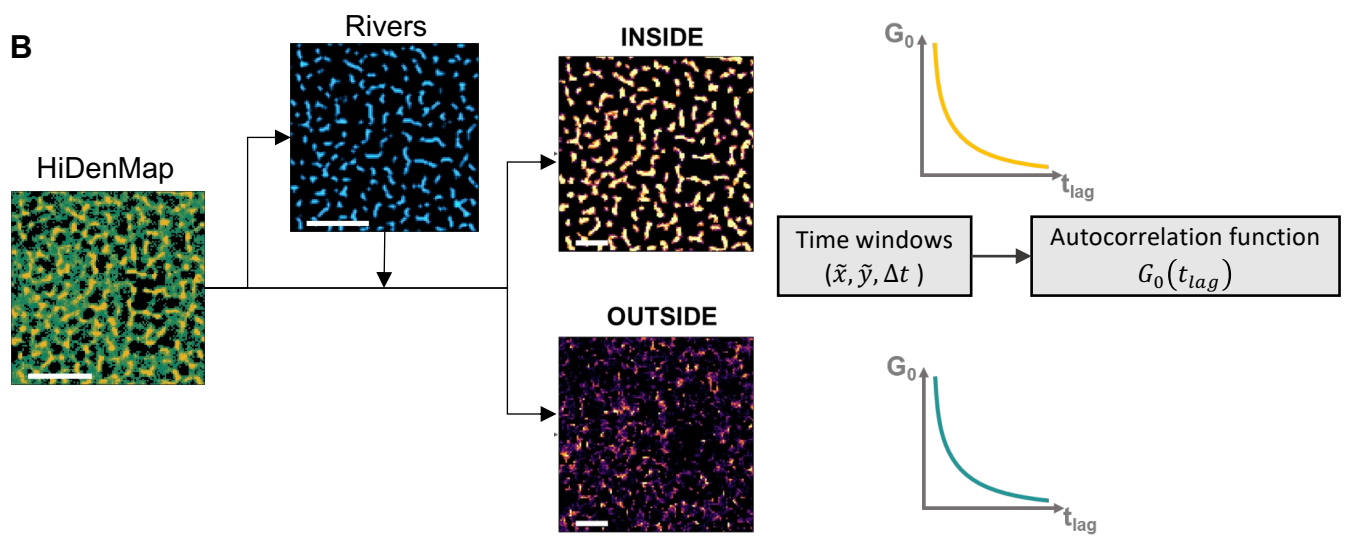

Figure 3

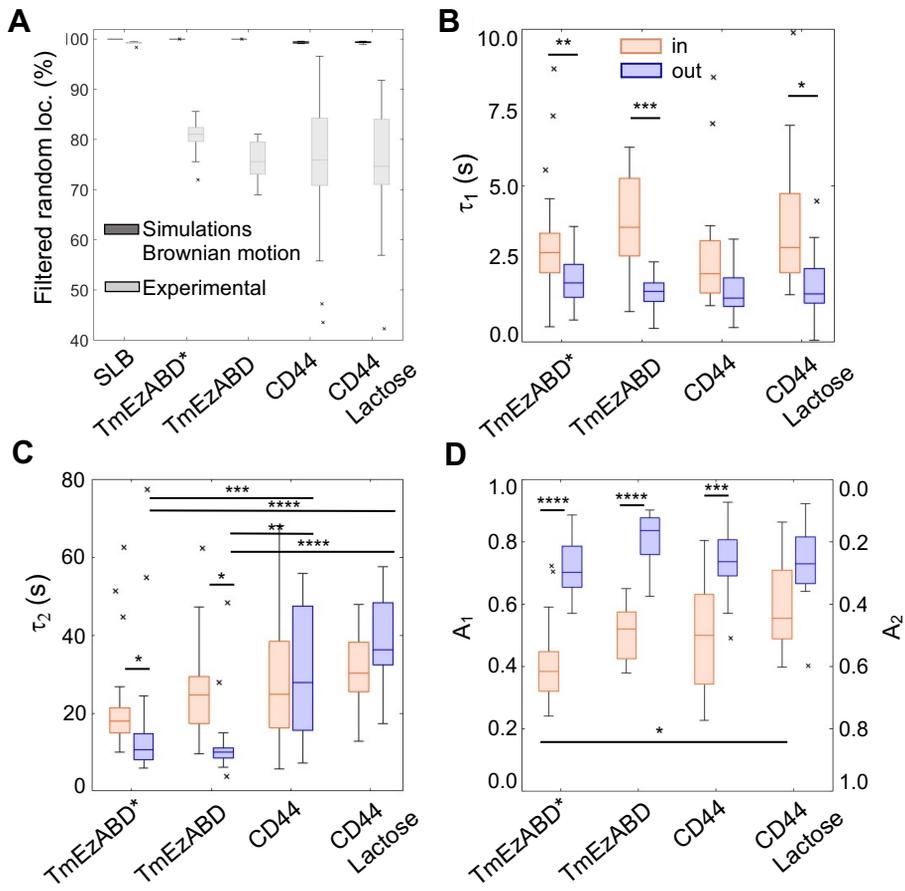

Figure 4

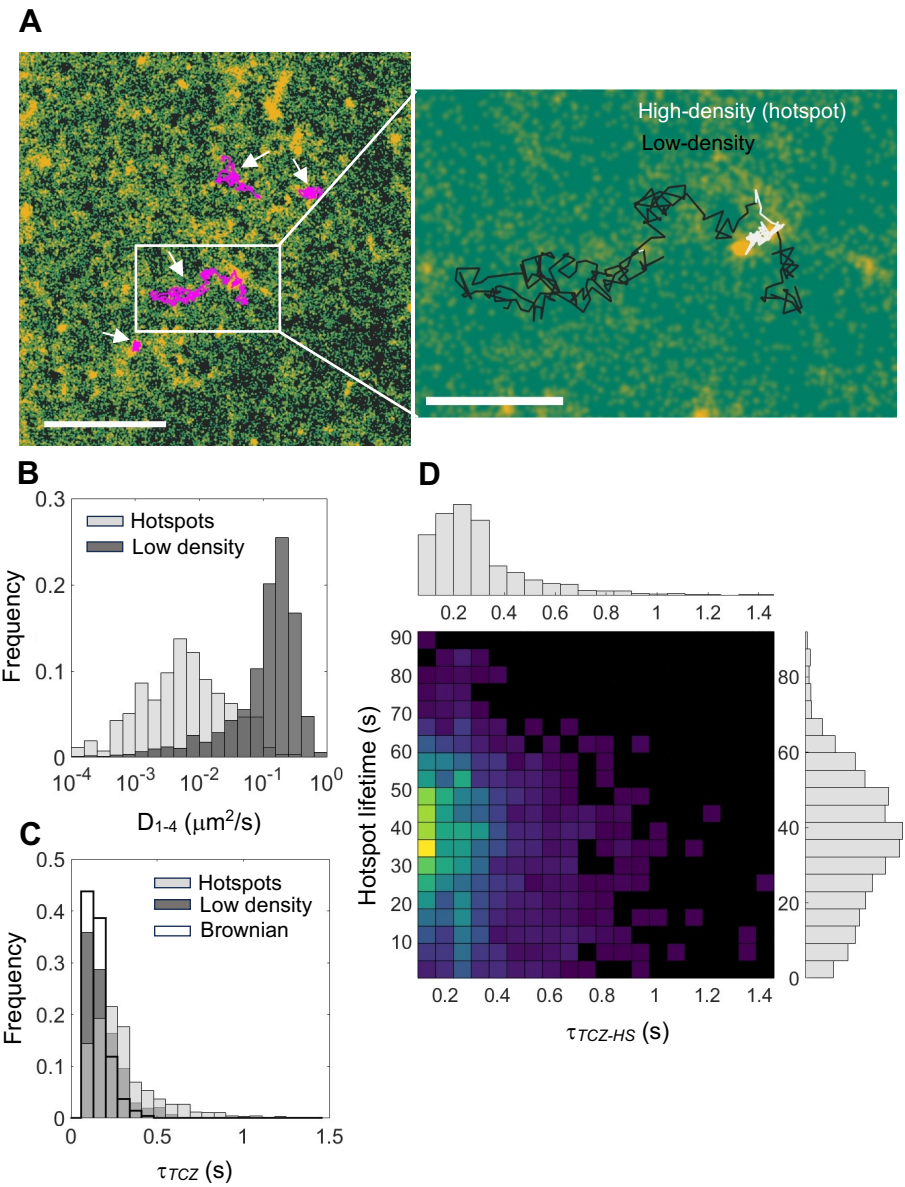

Figure 5